\documentclass[aps,prb,twocolumn,showpacs,floatfix,superscriptaddress]{revtex4}

\usepackage{amsmath}
\usepackage{amssymb}
\usepackage{epsfig}
\usepackage{graphicx}
\usepackage{dcolumn}
\usepackage{bm}

\begin{document}

\title{Stability of vortex structures in quantum dots}

\author{H. Saarikoski}
\email[ ]{henri.saarikoski@hut.fi}
\affiliation{Laboratory of Physics, Helsinki University of Technology,
P.O. Box 1100, FIN-02015 HUT, Finland}
\author{S.~M. Reimann}
\email[ ]{reimann@matfys.lth.se} 
\affiliation{Mathematical Physics, Lund Institute of Technology, 
SE-22100 Lund, Sweden}
\author{E. R\"as\"anen}
\affiliation{Laboratory of Physics, Helsinki University of Technology,
P.O. Box 1100, FIN-02015 HUT, Finland}
\author{A. Harju}
\affiliation{Laboratory of Physics, Helsinki University of Technology,
P.O. Box 1100, FIN-02015 HUT, Finland}
\author{M.~J. Puska}
\affiliation{Laboratory of Physics, Helsinki University of Technology,
P.O. Box 1100, FIN-02015 HUT, Finland}

\begin{abstract}
We study the stability and structure of vortices 
emerging in two-dimensional quantum dots in high magnetic fields.
Our results obtained with exact diagonalization and density-functional 
calculations show that vortex structures
can be found in various confining potentials.
In non-symmetric external potentials we find
off-electron vortices that are localized giving rise
to charge deficiency or holes in the electron density
with rotating currents around them. We discuss the role of quantum fluctuations
and show that vortex formation is observable in the energetics of the system.
Our findings suggest that vortices can be used to characterize the solutions
in high magnetic fields, giving insight into the underlying
internal structure of the electronic wave function.
\end{abstract}


\date{\today}

\maketitle

\section{Introduction}

Vortices can occur in quantum systems which are set to rotate, 
for example by applying an external magnetic field
or by mechanical rotation.~\cite{degennes,yarmchuk,butts} 
Using the Gross-Pitaevskii mean-field approach,
Butts and Rokhsar~\cite{butts} found that in a gas of 
rotating bosonic atoms which are weakly interacting with a repulsive force
between them, vortices may form in a crystal-like lattice,
in much analogy to patterns 
that emerge in rotating superfluid helium. These vortex solutions appear as 
holes (vortex lines) in the particle densities, where each single zero of the 
Gross-Pitaevskii wave function corresponds to a unit vortex. 
With increasing angular momentum, the bosonic cloud  
develops a flat shape, with more and more vortices penetrating it.

The analysis of the electronic structure of two-dimensional parabolic
quantum dots~\cite{jacak,reimannreview} (QD) 
in high magnetic fields has recently shown that, surprisingly,
vortices can appear also in {\it fermion} systems showing many similarities
to the boson case.~\cite{saarikoski,toreblad}
High-field solutions of the density-functional theory
revealed zeros in the electron densities, with electron currents circulating
around them. These were interpreted as vortex clusters.~\cite{saarikoski}
Many-body techniques could uncover vortex formation
at high magnetic fields\cite{saarikoski,tavernier}
or at large rotation\cite{toreblad}, giving credence to this interpretation.
With increasing magnetic field or rotation,  
successive transitions between stable vortex configurations were found.
These vortices carry magnetic flux quanta
similarly to the fractional quantum Hall effect (FQHE), where the
system can be approximated by Laughlin wave functions, which attach additional
vortex zeros at each electron.~\cite{fqhe} 
In quantum dots~\cite{saarikoski} and quantum dot molecules~\cite{ariprl}
vortices were found to be bound on the electron positions but,
unlike in the Laughlin wave functions, 
additional vortices appeared between the electrons.
A single vortex bound to an electron is a Pauli vortex, because it
is mandated by the exclusion principle.
In general, the number of vortices on top of each electron must be odd for
fermions to have correct particle statistics.
The additional vortices not bound to electrons
give rise to rotating currents of charge
and a charge deficiency down to zero electron density at the vortex centra.
By going around a vortex the wave function
gains a phase change of $2\pi l$, where $l$ is the winding number
or vortex multiplicity.

As characteristic for mean-field theories, 
the self-consistent solutions of the Kohn-Sham
equations for fermions, as well as the bosonic equivalent, the
Gross-Pitaevskii equations, break the symmetry of the quantum state.
For Bose-Einstein condensates under rotation, 
the Gross-Pitaevskii 
mean-field results have been shown to emerge as the correct leading-order 
approximation to exact calculations.~\cite{jackson2001} 
For Coulomb-interacting fermions, the density-functional 
approach suffers from the relatively crude approximations
for the exchange-correlation energy, as well as from the 
problems in using a single-configuration wave function.~\cite{harju2004}
It is thus important to compare the fermion mean field approach to a 
more rigorous solution of the full many-body Hamiltonian. 
However, this is a tedious enterprise: analytic solutions are out of reach 
for $N\ge 3$, and a numerically accurate direct diagonalization 
is also typically restricted to fairly small particle numbers.

For azimuthal symmetry of the confinement, 
the particle density retains this symmetry in the laboratory frame of 
reference. 
Therefore, to study the appearance of vortices in
those systems, a rotating frame,~\cite{maksym1996} 
conditional wave functions,~\cite{saarikoski}  
or correlation functions have to be examined.
In contrast to the bosonic case,~\cite{bosecor}
the pair correlations 
are not very informative for fermions due to the disturbing influence of the exchange hole.
Instead, the vortex solutions  
have been pinpointed either in a perturbative approach,~\cite{toreblad}
or using a conditional wave function which
fixes $N-1$ electrons to their most probable positions,
and then calculates the wave function of the $N$th 
electron.~\cite{saarikoski}
The latter method is not unproblematic either, since the
vortices are not independent of the electron dynamics. The vortex
locations in the conditional wave function depend on the positions of
the fixed electron coordinates as well as on the choice of
the probing electron.

In the FQHE the off-particle zeros are usually contrasted to on-particle
zeros.~\cite{murthy} The on-particle zeros are independent of electron
coordinates except the coordinate of the electron to which the vortex is
attached. On the other hand, the off-particle zeros of a given particle
are not necessarily off-particle zeros of other particles.
Graham {\em et al.} \cite{graham} used this fact to conclude that
off-particle zeros are not vortices in the real sense and therefore no
charge deficiency is necessarily associated with them.
Analogously to the FQHE case, the conditional wave functions
indicate that the electrons in quantum dots
see different positions for the off-particle
vortices, which is a manifestation of the fluctuations in the system.

In this paper we present direct theoretical evidence that in quantum
dots only the off-electron zeros show vortex structures in the
electron and current densities, and conclude that the vortex
structures are stable and can be used to classify the internal
structure of the many-electron wave function.
We do this by applying non-symmetric confining
potentials which leads to the localization of the vortices.
In the rotationally symmetric potential vortices are not localized
and they move as the electron coordinates are changed averaging
out the effect of vortices on the particle and current densities.
In rotationally {\it non-symmetric} potentials, however, the lower symmetry
should cause (at least partial) localization of vortices directly in
the particle density. 
The breaking of the circular symmetry was used 
already by Manninen {\em et al.} in the
study of Wigner localization of electrons in elliptical QD's
using exact diagonalization (ED) techniques.\cite{manninen}
Here we show that the same trick can be applied to vortices which become
directly visible in the exact particle and current densities.
We find, however, that the Pauli vortices at the
electron positions do not contribute to this effect.
In our calculations even a small asymmetry in the confining
potential is sufficient to localize vortices
and the fluctuations do not destroy this effect.
We compare the exact results to the mean field solutions and suggest that
the results can be generalized to arbitrary geometries.

\section{System characteristics}

We consider $N$ electrons trapped by a confining potential $V_c$
and subject to a perpendicular, homogeneous magnetic field ${\bf B}=(0,0,B)$. 
The system is described by an (effective-mass) Hamiltonian
\begin{equation}
H=\sum^N_{i=1}\left[\frac{({\mathbf p}_i+e {\bf A})^2}{2m^*}
+V_{\rm c}({\bf r}_i)\right] + \frac{e^2}{4\pi \epsilon} \sum_{i<j}
\frac{1}{\mid {\bf r}_i - {\bf r}_j \mid } \ ,
\label{hamiltonian}
\end{equation}
where ${\bf A}$ is the
vector potential of the magnetic field $B$,
$m^*$ the effective electron mass, and $\epsilon$ is the dielectric
constant of the medium. We apply the typical material parameters for GaAs, 
namely, $m^*/m_e=0.067$ and $\epsilon/\epsilon_0= 12.4$.
We give the energies and lengths in effective atomic units,
i.e., in Ha$^*\simeq 11.86$ meV and in $a_B^*\simeq 9.79$ nm.

At high magnetic fields, after complete polarization of the QD,
the exchange energy results in the formation of a stable and compact
structure, the so called maximum density droplet (MDD).\cite{macdonald}
It is a finite-size equivalent of the $\nu=1$ quantum Hall state
which assigns one Pauli vortex at each electron position.
The MDD state can be found in various QD geometries.
In circularly confined QD's the electrons occupy successive
angular momentum states on the lowest Landau level, which
in the parabolic case leads to a relatively flat electron density.
In rotationally non-symmetric potential wells, instead,
the MDD window can be deduced from the kinks in the chemical
potentials.
In addition, the magnetic field for the MDD
formation can be accurately predicted from the number of flux quanta
penetrating the QD.~\cite{Esa}

As the magnetic field is increased, the compact electron droplet 
is squeezed and eventually the repulsive interactions between the 
fermions cause the MDD to reconstruct. 
For parabolic QD's, different scenarios
of the reconstruction have been suggested.
Chamon and Wen\cite{chamon} found a ``stripe phase'' 
where a lump of electrons separates from the MDD at a distance 
$\approx 2\ell _B$, where $\ell _B = \sqrt{\hbar /eB}$  is the 
magnetic length.
Goldmann and Renn introduced projected necklace states
which they found to be lower in energy than the states found by Chamon and
Wen.~\cite{goldmann}
Geometrically unrestricted  Hartree-Fock\cite{koonin}
and CSDFT\cite{reimann99,saarikoskiphysicae} studies suggested that such 
edge reconstruction would occur with a modulated charge 
density wave along the edge.
For a sufficiently small Zeeman gap, this polarized reconstruction
may be preempted by edge spin textures.~\cite{karlhede,sami}

The ED shows instability with respect to addition of internal holes
as discussed by Yang and MacDonald.~\cite{yang}
These holes were recently 
re-interpreted as vortices in mean-field density-functional
calculations.~\cite{saarikoski}
They were found also further away from the dot center
where the electron density is low.
Vortices behave often like classical (localized) particles in the mean-field
approach and the vortices appear as rotating currents of charge with a zero
in the particle density at the vortex centra.
Several charge-density-wave states that mix different eigenstates
can also be interpreted as solutions describing a transport of a vortex
to the center of a QD.~\cite{saarikoskiphysicae}
Formation of vortices causes usually a broken symmetry in the 
mean-field particle density, even when the Hamiltonian is 
azimuthally symmetric.

Tavernier {\it et al.}~\cite{tavernier} studied distribution of zeros
in the exact many-body wave function of systems containing
up to 4 electrons.
They compared the results with the rotating-electron-molecule (REM)
wave functions.~\cite{yannouleas1}
In the regime where the effect of the external confining potential
can be neglected, the rotating electron molecule model can provide
an intuitive description of the Wigner-localized electrons.
Tavernier and coworkers\cite{tavernier}
found out, however, that the REM model is unable 
to predict the clustering of vortices near electrons.

\section{Numerical results}

\subsection{Mean-field description of vortex solutions}

In order to solve the many-body 
Schr\"odinger equation corresponding to the Hamiltonian 
(\ref{hamiltonian}), we first work in the mean-field picture and apply 
the spin-density-functional theory (SDFT).
For the self-consistent solution of the Kohn-Sham equations we employ
a real-space scheme,~\cite{saarikoski1} where the external confining potential
$V_{{\rm c}}$ can be arbitrarily chosen without symmetry restrictions.
The exchange-correlation effects are taken into account using the local
spin-density approximation (LSDA).~\cite{LDA}
At high magnetic fields, the effect of currents in the 
exchange-correlation potentials becomes non-negligible, and the
current-spin-density-functional theory (CSDFT)\cite{vignalerasolt}
gives a slightly better approximation to the ground state energy.
\cite{saarikoskiprb}
The CSDFT is computationally more demanding than the SDFT but
according to our test calculations qualitatively similar
vortex structures were found to emerge in both formalisms.
We apply the SDFT throughout this paper, since we found in these tests
that it captures all the essential physics of these systems at much lower
computational work.

The confinement is chosen to be a two-dimensional 
harmonic oscillator potential with elliptic deformation, 
defined as
\begin{equation}
V_{{\rm c}}({\bf r})=
{1 \over 2} \hbar \omega_0^2(\delta x^2+{1\over \delta} y^2),
\label{Vell}
\end{equation}
where $x$ and $y$ are the major axes of the ellipse, $\hbar \omega$
is the confinement strength, and $\delta$ is the eccentricity.

\begin{figure*}
\begin{center}
\begin{minipage}{0.68\linewidth}
\epsfig{file=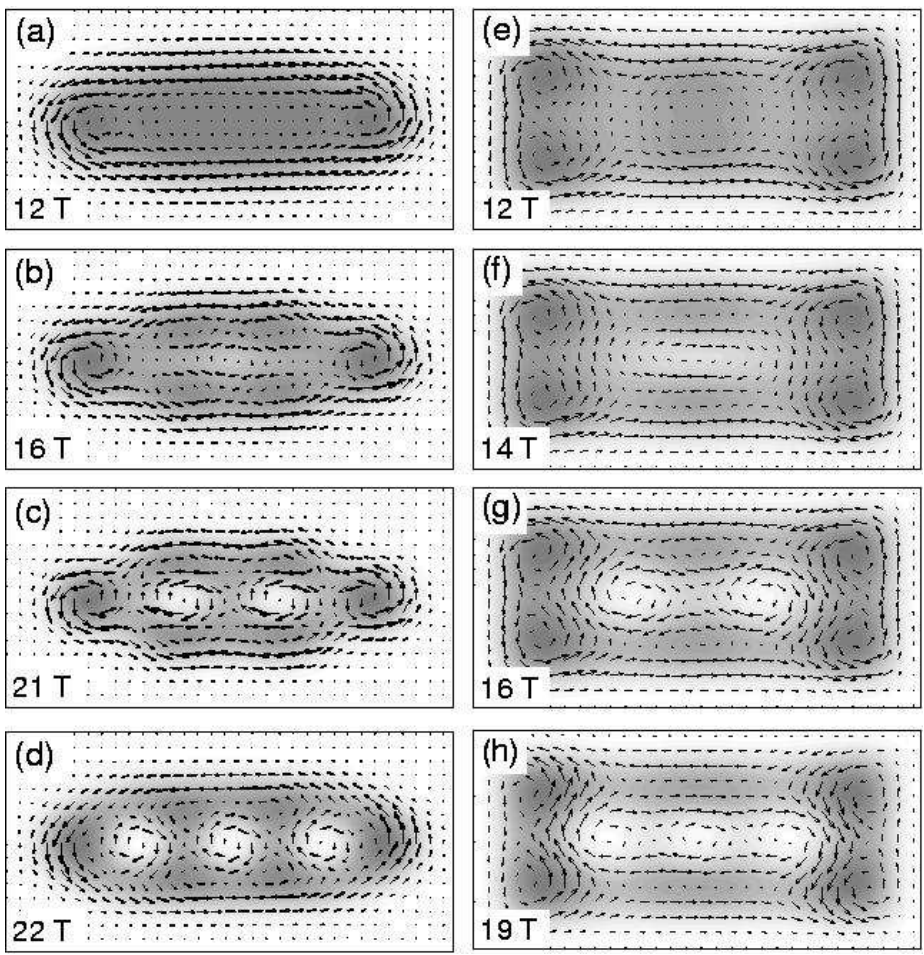, width=\linewidth}
\end{minipage}\hfill
\begin{minipage}{0.30\linewidth}
\caption{Spin-density-functional-theory (SDFT)
electron densities and currents in
an elliptic (a-d) and rectangular (e-h) six-electron 
quantum dot in different magnetic fields.
The potential parameters in effective a.u. are 
($\hbar \omega,\delta$)=($0.5,2$) and
($L_x,L_y$)=($2\sqrt{2}\pi,\sqrt{2}\pi$) in elliptic and rectangular
dots, respectively.
The increasing of the magnetic field leads
to a formation of a vortex pattern beyond
the maximum density droplet (MDD)  solution ($B=12$ T)
in both geometries.\label{fig:sdftrectangle}}
\end{minipage}
\end{center}
\end{figure*}

For comparison, we also apply a rectangular hard-wall 
confinement, defined in Ref.~\onlinecite{Esa}. The parameter describing 
the deformation of the confining potential in this case is
the side-length ratio $\beta=L_x/L_y$.
In the post-MDD domain, the SDFT predicts the formation of 
vortices inside the QD's. This is visualized in 
Fig.~\ref{fig:sdftrectangle}
showing solutions containing up to three vortices 
in elliptic (a-d) and rectangular (e-h) six-electron QD's.
The eccentricity $\delta$ and the side-length ratio $\beta$ have been set 
to $2$ in these systems, respectively. 
As the vortices repel each other, the MDD states shown in 
Figs.~\ref{fig:sdftrectangle}(a) and (e) reconstruct into
states that enclose a linear vortex pattern along the longest
major axis. There is a remarkable qualitative 
similarity in the high magnetic field behavior of these systems. 
However, a linear vortex cluster requires a rather
large eccentricity (side-length ratio) of the QD. 
In the elliptic case with a confinement strength of  
$\hbar\omega =0.5 $ Ha$^*$, the triple-vortex
configuration changes from triangular to linear when 
$\delta$ is increased to about 1.4.

Figure~\ref{fig:sdftrectangle} shows currents induced by the
magnetic field. The current is flowing clockwise
around the vortices and anticlockwise on the edges of the dot.
The reversal of the current near the vortex core
is due to inner circulation of the electrons.\cite{lent}
The number of vortices in the QD increases with the magnetic field,
and the current loops of the vortices start to overlap.
This causes formation of giant current loops which comprise several vortices.

\begin{figure}
\epsfig{file=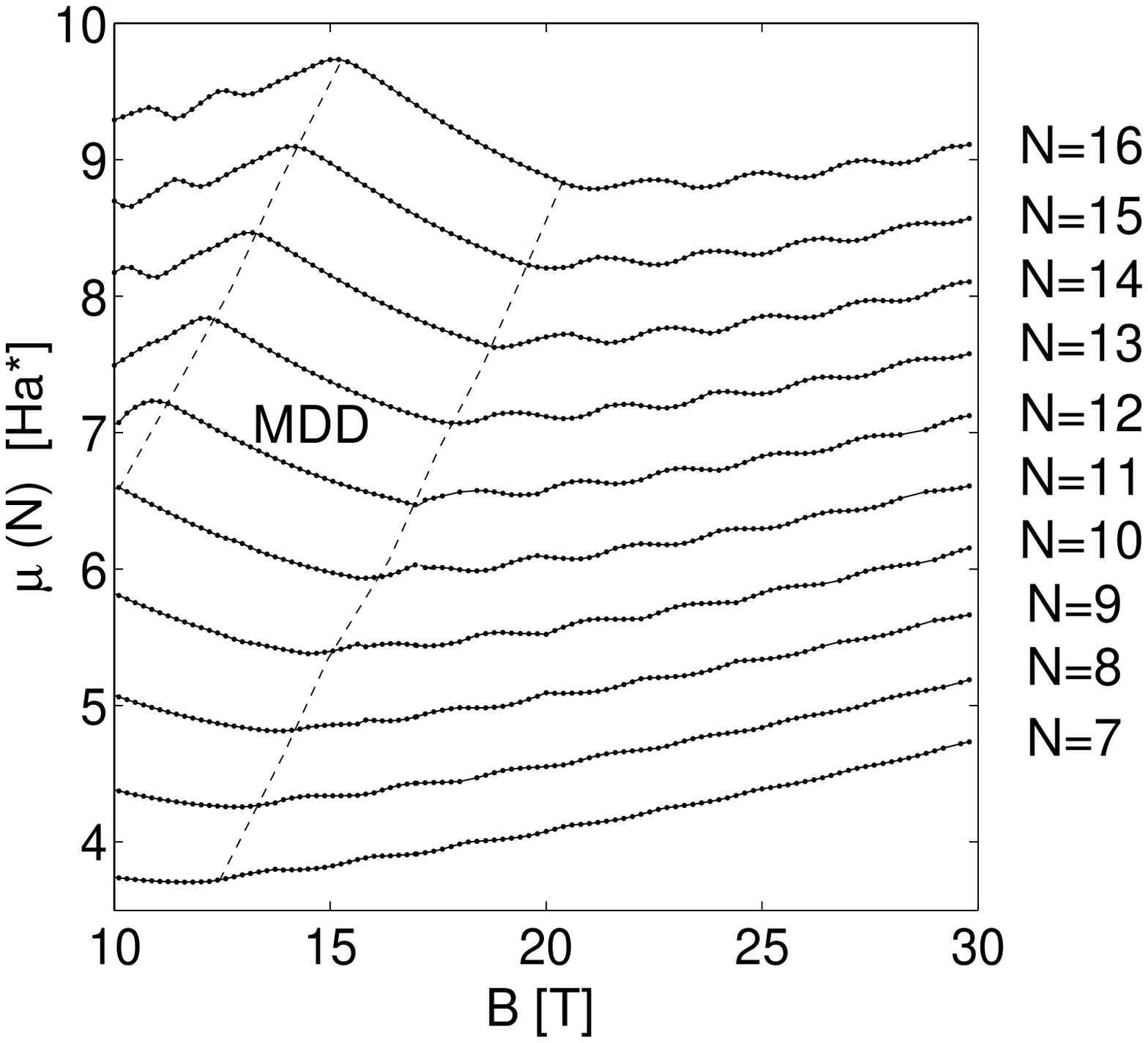, width=8cm}
\caption{Chemical potentials for $N$-electron rectangular
($\beta=2$) quantum dots as a function of the magnetic field.
The dot is expected to be fully spin-polarized.
After the MDD window there are oscillations
in the chemical potential which matches with the appearance of additional
vortices (one-by-one) into the quantum dot.
\label{chem}}
\end{figure}

\begin{figure}
\epsfig{file=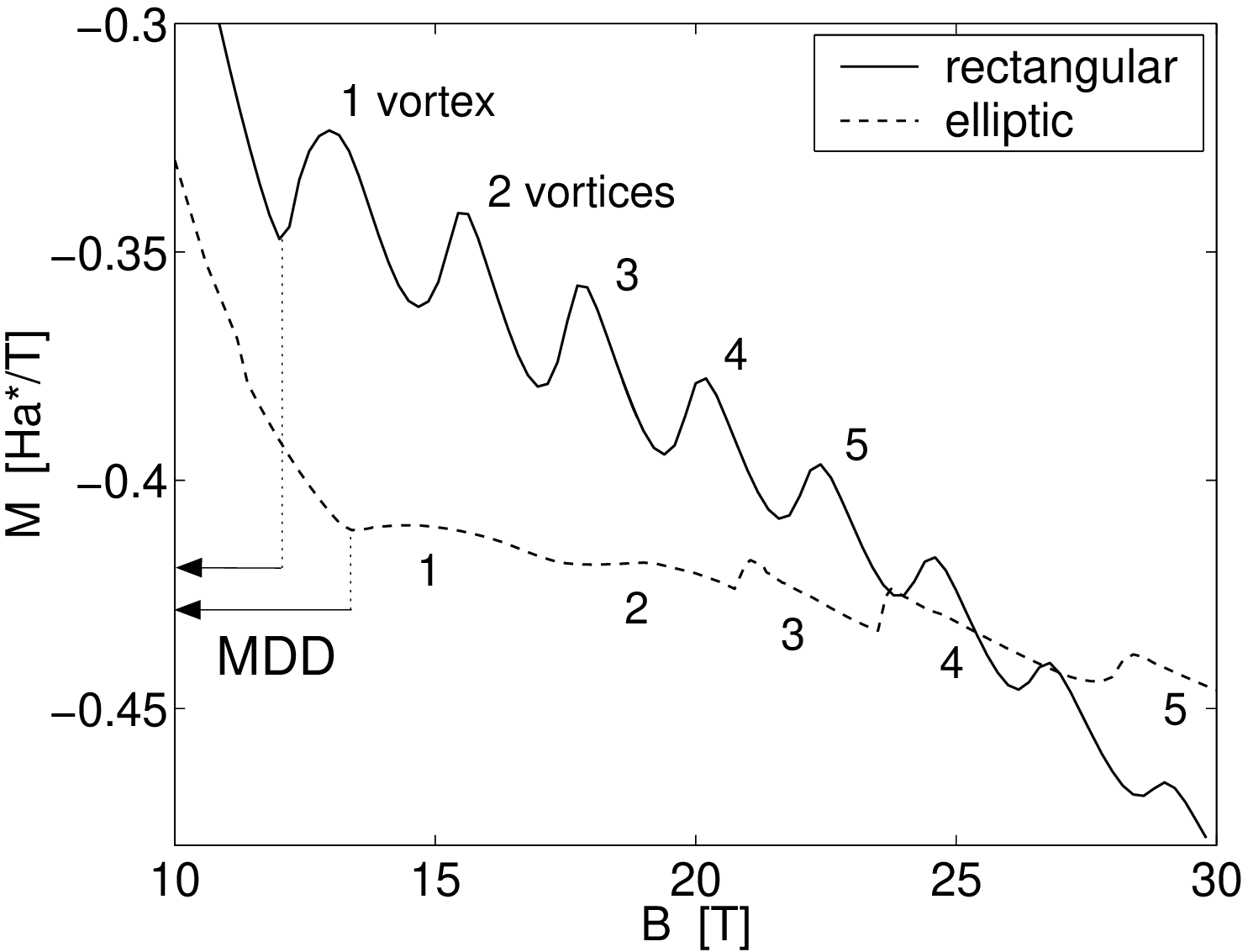, width=8cm}
\caption{Magnetization of a rectangular (solid line) and
elliptical (dashed line) six-electron quantum dot as a function of
the magnetic field. The ellipse eccentricity $\delta$ and the rectangle
side-length ratio $\beta$ are both set to 2.
The numbers in the figure denote the correponding number of vortex
holes in the electronic structure.
\label{magN6}}
\end{figure}

Fig.~\ref{chem} shows the chemical potentials,
$\mu(N)=E(N)-E(N-1)$, of a rectangular ($\beta=2$) QD 
containing $N=7\ldots 16$ electrons. 
Interestingly, the regime beyond the MDD is characterized by periodic
oscillations in $\mu$ as a function of the magnetic field $B$. The
peak positions in the oscillations match with the transitions between
adjacent vortex states and mark the emergence of additional vortices
one-by-one in the QD. This is presented for $N=6$ in
Figs.~\ref{fig:sdftrectangle}(f)-(h). Note that 
the oscillations get stronger as the number of electrons increases.
The origin of the oscillations lies in the large
reduction of the Coulomb energy in connection with
the vortex formation and the coexistent pronounced
localization of the electrons. The oscillations are also visible in the
total magnetization,
$M=-{\partial E_{\rm tot}}/{\partial B}$, as shown in Fig. 4 in
Ref.~\onlinecite{Esa}. The magnetization in rectangular and
elliptic six-electron QD's are compared in Fig. \ref{magN6}.
In elliptic QD's the oscillations are weaker and less regular than
in their rectangular counterparts where the size of the dot is a
constant. The soft-wall confinement makes the dot more flexible in
minimizing the total energy. These results are in accord with
the exact diagonalization results by Goldmann and Renn~\cite{goldmann}, who
calculated the chemical potentials of QD's in parabolic
and non-parabolic ``coffee-cup'' shaped confinements.
The latter confinement has a hard wall similarly to
our rectangular potential well.
Goldmann and Renn found kinks in the chemical potentials
and the kink sizes increase with the electron number.
Moreover, the kinks in the non-parabolically confined systems
were found to be much larger than in the parabolically confined systems.

The results above suggest that the vortex
formation is a considerable energetic effect that could be detected in
appropriate experiments.
Oosterkamp {\em et al.}\cite{oosterkamp}
measured the Coulomb oscillations peaks (chemical potentials) in 
transport experiments for vertical QD's and observed
additional phase transitions beyond the MDD. 
They found oscillations in chemical potentials and
the amplitude of these oscillations increased with the electron number.
These data are consistent with our calculations and could indicate
vortex formation in quantum dots.
A direct interpretation
of their result is, however, difficult due to the
unknown shape of the QD sample
and its eventual sensitivity to disturbance in the experiment. 
Moreover, the experimental oscillations may also indicate 
other phenomena, such as the formation of a spin texture,
for example.~\cite{karlhede}

\begin{figure}
\epsfig{file=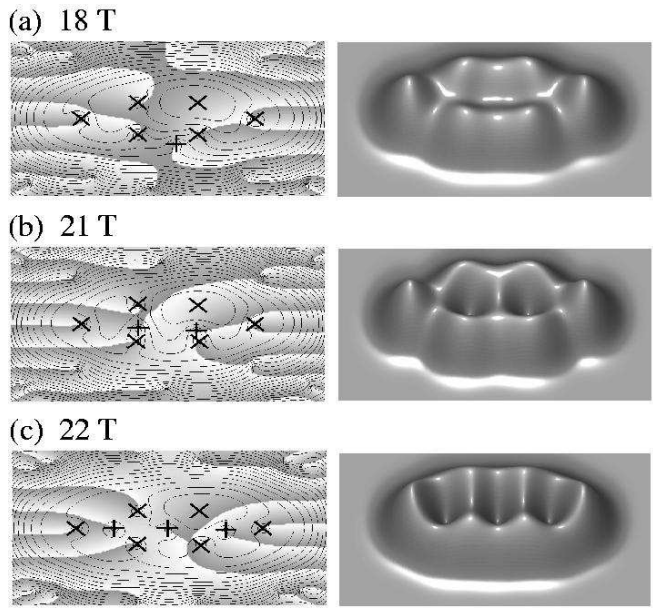, width=8.6cm}
\caption{SDFT solutions of six-electron, elliptically ($\delta=2$) confined 
quantum dots
at different magnetic fields. The confinement strength is set to 
$\hbar \omega = 0.5$ Ha$^*$. 
{\em Left panel}: Conditional single-determinant wave functions of the
Kohn-Sham states. The fixed electrons
are marked with crosses and the probing electron is the rightmost
electron at the top. The contours show the logarithmic electron density
of the probe electron and the grey-scale show the phase of the wave function.
The phase changes from $\pi$ to $-\pi$ at the lines where shadowing
changes from the darkest grey to white. The vortices that cause
charge deficiency in the center of the dot are marked with + signs.
{\em Right panel}: Electron densities show vortex holes
and partial localization of the electrons forming six density maxima.}
\label{ksdet}
\end{figure}

Conditional wave functions can be introduced not only for the analysis 
of the exact many-body wave function, as described in 
Ref.~\onlinecite{saarikoski}, but they are also useful to analyze the SDFT
results. We use an auxiliary single-determinant function
of the Kohn-Sham orbitals which emulates
the exact conditional wave function.~\cite{saarikoskiphysicae}
This allows a study of the SDFT
solutions where the electron density may have several minima,
but the vortices are not directly localized to fixed positions. This may be
due to the mixing of several eigenstates as shown in
Ref.~\onlinecite{saarikoskiphysicae}.
For instance, the electron densities for the SDFT states at 18 T and 21 T 
show no density zeros. However, the corresponding conditional
single-determinant functions show vortices near
the fixed electron ring (see Fig.~\ref{ksdet}).
In this picture Pauli vortices can be seen on the fixed electrons, and
additional vortices are found between the electrons. 

The electron densities of the elliptical $\delta=2$ dot
in Fig. \ref{ksdet}
show localization of both electrons and vortices.
The electron localization results in six density maxima which is consistent
with earlier SDFT calculations of elliptically
confined QD's.~\cite{manninen} 
The additional vortices give rise to charge deficiency in the electron density 
as seen in the right panel of Fig. \ref{ksdet}.
According to the SDFT results, the intensity of the vortex localization 
strongly depends on the magnetic field. In the above example, the
localization of vortices is partial with pronounced minima in the
electron density in solutions of up to two vortices.
As the magnetic field is increased further, the localization becomes complete,
{\em i.e.}, the electron density vanishes at the vortex core.
This reflects the decrease of the magnetic length $\ell_B$.

The results indicate that only the additional vortices between the electrons
have a charge deficiency associated with them.
Conditional wave functions\cite{saarikoski}
and the total electron densities calculated with the SDFT
show that there are vortices also further away from the dot center.
In this region the electron density is usually a tiny fraction of
the maximum electron density. The quantum fluctuations smooth out the
effect of these external vortices and they cannot be directly observed
in the exact particle density, even though the rotational symmetry is broken.

\subsection{Exact diagonalization for an elliptic dot}

We compare now the above results obtained within SDFT to those of
a direct numerical diagonalization of the many-body Hamiltonian matrix. 
We hereby focus on small particle numbers and magnetic fields high enough such 
that the description is to a large extent restricted to what in the 
isotropic case would correspond to the lowest Landau level. 
In order to display the
internal structure of the many-body wave function, we 
break the spherical symmetry of the dot 
by applying the elliptical confining potential [Eq.~(\ref{Vell})]. 
We assume full polarization of the electron droplet, and neglect the 
Zeeman energy.

For the deformed case, $\delta\not= 1$, where the total angular
momentum is not any more a good quantum number (while, however, we
still have good parity), the most appropriate and efficient basis
set spanning the Fock space is formed by the eigenstates to the
single-particle part of the Hamiltonian Eq.~(\ref{hamiltonian}).
These must be calculated numerically. We determine the 
$\cal M$ lowest ones by directly diagonalizing the single-particle part
of the Hamiltonian in a basis consisting of 
a sufficient number of corresponding Fock-Darwin~\cite{fockdarwin} states
at $\delta=1$, which are known analytically.
Once the single-particle basis is at disposal, the Fock states are generated
by sampling over all possibilities to 
set $N$ particles on these $\cal M$ states.
From this sampling, only those Fock states with defined parity
and configuration energy~\cite{wignerpap} 
less than a defined cut-off energy, are chosen for diagonalization.
The cut-off energy was adjusted to restrict the number of 
Fock states (i.e. the matrix dimension) to be less than about 50000.
We limit the single-particle basis dimension to ${\cal M}\leq 44$ 
and use numerical integration for calculating the Coulomb matrix
elements. 
The many-body Hamiltonian (\ref{hamiltonian}) is then
diagonalized in the obtained subspace.
Densities $\langle\hat n({\bf r})\rangle$ and real currents 
$\langle\hat {\bf j}({\bf r})\rangle$
are finally calculated in order to compare the broken-symmetry 
solutions of the ED directly to the mean-field
results.
Here, 
\begin{equation}
\hat n({\bf r})=\sum_i\delta({\bf r}-{\bf r}_i)
\label{density}
\end{equation}
is the density operator. The real current is obtained by taking the 
expectation value of 
\begin{equation}
\hat{\bf j}({\bf r})=\hat {\bf j}_p({\bf r})+
{{e}\over{m^*}}{\bf A}({\bf r})\hat n({\bf r})~, 
\label{current}
\end{equation}
where 
\begin{equation}
\hat{\bf j}_p({\bf r})=\sum_i{{-i\hbar}\over{2m^*}}\left[
\delta({\bf r}-{\bf r}_i)\nabla_i+\nabla_i\delta({\bf r}-{\bf r}_i)\right]
\label{pcurrent}
\end{equation}
is the paramagnetic current operator. 

We should note at this point that for obtaining an accurate description 
of the total energy of the system, using only 44 lowest single-particle states 
is not sufficient for a full convergence of
the total energy. However, the relative energy differences and 
the geometrical structure of the electron and current densities of the
ground-state and lowest-lying states were converged within this basis set.
Since our aim is the comparison of the broken-symmetry many-body 
with the mean-field solutions, rather than a detailed discussion of energy spectra and
excitation energies, this truncation appeared reasonable. 

\begin{figure}
\hbox{\epsfig{file=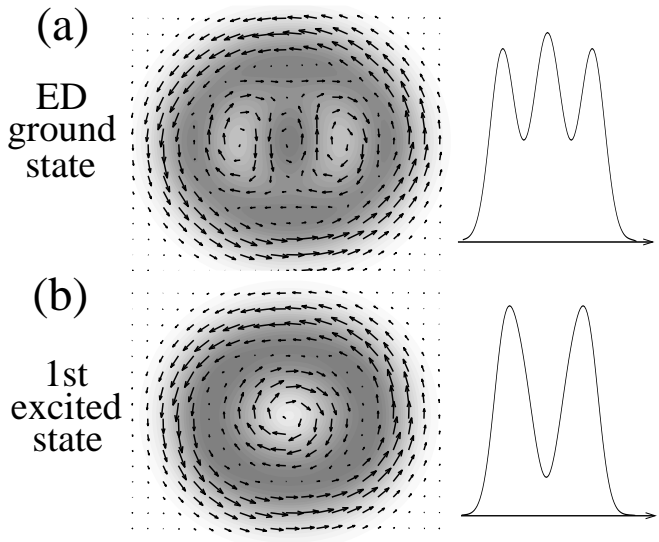, width=\columnwidth}}
\caption{
Electron densities (grey scale) and current densities (arrows)
of exact diagonalization (ED) solutions for an elliptical quantum dot
with $\delta=1.1$. 
(a) The ground state with a two-vortex structure. 
(b) The first excited state with a single vortex
at the origin.  
The right column shows the corresponding electron densities at
the longest major axis of the ellipse.
The confinement strength is $\hbar \omega=0.5$ Ha$^*$ 
and the magnetic field is $B=13.4$ T. 
}
\label{fig:delta1.1}
\end{figure}

Figure~\ref{fig:delta1.1} 
shows the electron and current densities, 
$\langle \hat n({\bf r}) \rangle $  and  
$\langle\hat {\bf j}({\bf r})\rangle$, of the ground state and first excited
state for an elliptic dot with eccentricity 
$\delta = 1.1$ and magnetic field 13.4 T. 
Both states have parity $\pi = -1$, and the first excited
state is separated from the ground state by only 7.7 mHa$^*$. 
The ground state shows a vortex pattern around two minima in the density 
and its total angular momentum is $L=-25.4\hbar $. 
The leading single-particle configuration~\cite{toreblad}  
of the Fock state has the form $\mid 1001111100...\rangle $, with amplitude  
$\mid c_{\cal  L}\mid ^2= 0.4$.
The first excited state with $L=-21.4\hbar $ shows a pronounced 
single vortex at the dot center, with the current circulating clockwise around
the origin. The hole shows the characteristic cone shape for a vortex and  
appears nearly localized at the dot center. The leading 
configuration of the Fock states is $\mid 011111100...\rangle $
with amplitude $\mid c_{\cal  L}\mid ^2= 0.7$.
The second excited state has parity $\pi = +1$ and is separated by a gap of 
53 mHa$^*$ from the ground state. It has angular momentum $L=-20.4\hbar $ 
and no clear vortex structure. 

\begin{figure}
\hbox{\epsfig{file=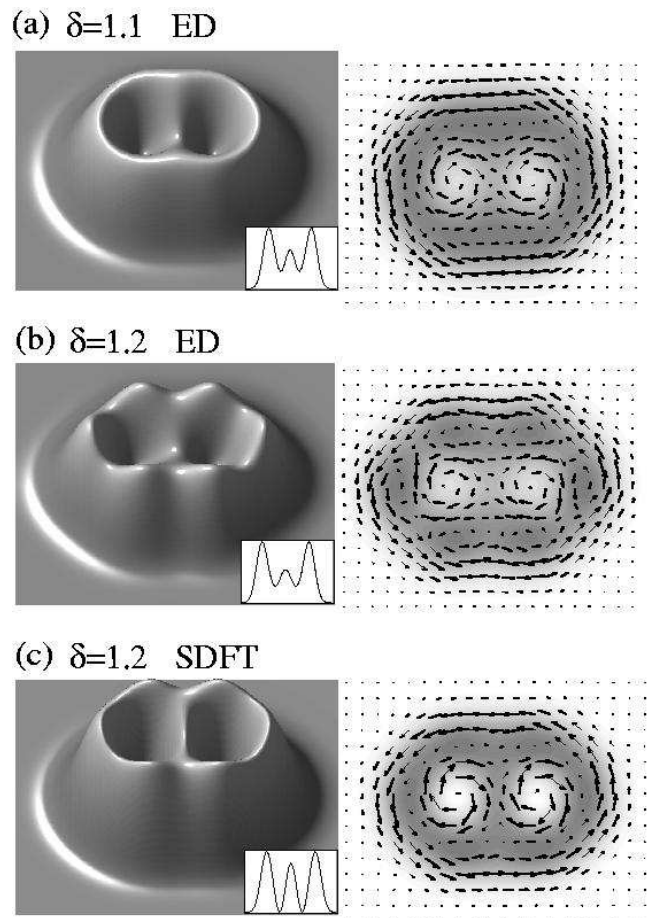, width=8.5cm}}
\caption{Electron densities (grey scale) and current densities (arrows)
for two-vortex solutions in elliptical QD's. 
(a) Exact diagonalization (ED)
result at $\delta=1.1$. (b) ED result at $\delta=1.2$.
(c) SDFT result at $\delta=1.2$.
The inset shows the corresponding electron densities at
the longest major axis of the ellipse. The vortices
are localized in the SDFT solution, and the ED
solution shows slight delocalization due to quantum fluctuations.
The confinement 
strength is $\hbar \omega=0.5$ Ha$^*$
and the magnetic field is $B=17$ T.
}
\label{fig:delta1.2}
\end{figure}

Figure~\ref{fig:delta1.2} 
shows the densities and density contours for 
the cases $\delta =1.1$ (a) and $\delta =1.2$ 
(b) at an increased magnetic field of 
$17$~T. Both states have parity $\pi =-1$ and angular 
momentum $L=-26.7\hbar $ (a) and  $L=-28.5\hbar $ (b).
In both cases, the current circulates around fairly  
pronounced minima in the charge density, 
representing two vortices at the dot center. 
When the eccentricity increases from $\delta =1.1$ (a) to $\delta =1.2$ (b), 
the electrons begin to arrange themselves in the form of a Wigner molecule
with counterclockwise rotation of the current around the ${\it maxima }$ of
the electron density, and clockwise rotation around the vortices 
(i.e. the minima in the density), as it was 
also observed at $\delta = 1.1$.  The reversal of the current near
the vortex core is consistent with the SDFT results.

It is important to note here that the ground state in both cases is a state 
with $\pi = +1$, which for (a) is 34 mHa$^*$ 
and for (b) is 7.6 mHa$^*$ lower in energy.
No states were found in between these two lowest states with $\pi =+1$
and $\pi = -1$, i.e. both states shown in (a) and (b)  
appeared as first excited states, respectively.
The fact that the energetic sequence of the states changes with increasing 
field is, however, not surprising: 
It is well known\cite{reimannreview} 
from the ED studies of circularly symmetric QD's, that 
the {\it overall} structure of the spectrum is 
independent of the magnetic field, the role of which is mainly to tilt the
spectrum so that the minimum energy is at a different state
(see Manninen {\it et al.}\cite{manninen}, and 
Maksym and Chakraborty\cite{maksym}). 
The above results from ED are compared 
with the mean-field calculation in Fig.~\ref{fig:delta1.2} (c) for 
$\delta =1.2$. 
Due to the approximations made in SDFT, the transitions between
different vortex solutions appear at slightly
different $B$-values. This fact, together with the tilt of the ED spectrum
with increased $B$, can explain why in Fig.~\ref{fig:delta1.2}
the first excited state in the ED solution compares to the
ground-state two-fold vortex obtained by SDFT.
In the ED result, quantum fluctuations destroy the complete
localization of the vortices, i.e. the exact density 
at the vortex center still is about one third of the maximum density, 
while in the mean-field result the density at the vortex center is reduced to
zero. 

\begin{figure}
\hbox{\epsfig{file=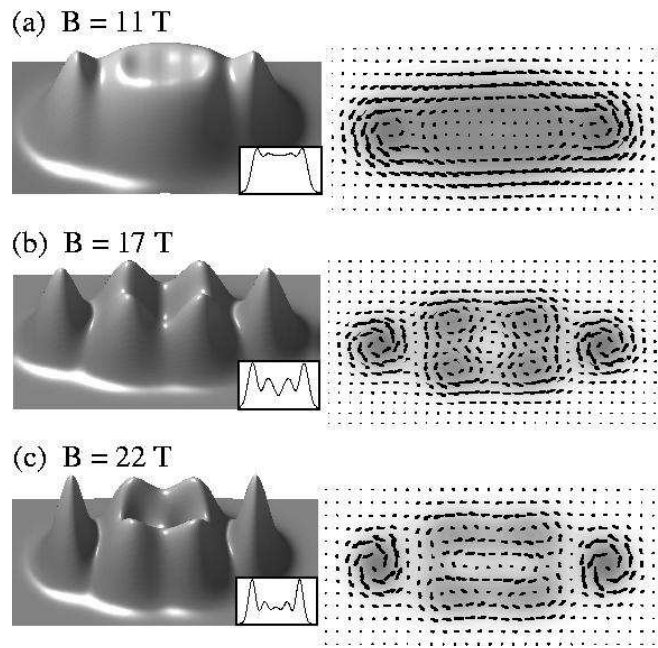, width=8.7cm}}
\caption{
Electron densities (grey scale) and current densities (arrows)
of ED solutions for elliptical quantum dots with $\delta=2$.
The magnetic field is (a) 17 T, (b) 21 T, and (c) 22 T.
The inset shows the corresponding electron densities at
the longest major axis of the ellipse.
The confinement strength is $\hbar \omega=0.5$~Ha$^*$.
}
\label{fig:eddens}
\end{figure}

Increasing the eccentricity still further, the localization of electrons 
leads to the formation of a 
charge-density-wave-like crystal. Vortices in between the 
classical electron positions  become even more apparent. 
Figure~\ref{fig:eddens} 
shows the 
ED electron densities (grey scale) and current densities (arrows) for larger  
eccentricity $\delta=2$. The magnetic 
field is (a) 11 T,  (b) 17 T and (c) 22 T, and $\pi =-1$ 
in all cases. (The $\pi =+1$ states are slightly higher in energy).  
The solution at 11 T shows the MDD, and vortices form at higher magnetic
fields. These solutions can be compared to the
SDFT solutions in Fig.~\ref{fig:sdftrectangle}.
There is again a qualitative agreement, but now the ED shows stronger
localization of electrons at both ends of the dot.
While now the Fock states show much mixing and a clear dominance of a
few single configurations 
could not be observed, the general trend is very similar to the SDFT result
shown above in Fig.~\ref{fig:sdftrectangle}.

\section{summary}

We have found that in non-circular confinements
the vortices can be directly seen in the exact diagonalization electron 
densities as holes or charge deficiencies around which the current 
is circulating.
The vortices are stable against the quantum fluctuations and
even a slight asymmetry in the confining potential cause the vortices
to show up in the exact many-body electron density as density minima.
However, due to quantum fluctuations, there are no zeros in the
electron density.
The spin-density-functional calculations are in accordance with these results.
They predict analogous vortex formation
also in rectangular hard-wall quantum dots suggesting that the results
can be generalized to a wide variety of geometries.
The chemical potentials of the rectangular dot show features in
the energetics which suggest the possibility of a direct comparison to 
experiments.

In the light of these results it is justified to speak of vortices as real
quasi-particles.
Vortices are not independent of the electron dynamics, but
they can be used to characterize the solutions in
high magnetic fields which gives insight of the underlying internal structure
of the electronic wave function.
Prediction of vortex formation in hard-wall potentials gives also credence
to the assumption that the vortices are robust and largely independent
of the chosen geometry of the system. Therefore they seem to be universal
features of the physics of the two-dimensional interacting fermion systems
in strong magnetic fields above the maximum density droplet formation. 

To conclude, we briefly compare these results to {\it bosonic} systems, where
vortex formation in rotating Bose-Einstein condensates has been much discussed 
both theoretically and experimentally. There are apparent similarities between 
the bosonic and fermionic case\cite{toreblad}.
In the composite fermion model bosons can be turned into fermions
(and the other way around) by attaching fictious magnetic fluxes on top
of the electrons. Vortex formation appears as a universal phenomenon
of quasi-two-dimensional quantum systems suggesting
a synthetic theoretical rationale behind the phenomena.~\cite{iggi}

\acknowledgments

Special thanks are to Matti Koskinen for his help with the exact
diagonalization code. We also thank Matti Manninen,
Ben Mottelson, and Maria Toreblad for fruitful discussions. 
This work has been supported by the Academy of Finland through the Center
of Excellence Program (2000-2005), the Swedish Foundation for Strategic
Research (SSF) and the Swedish Research Council (VR).


\end{document}